# First-principles study of intrinsic and hydrogen point defects in the earth-abundant photovoltaic absorber $Zn_3P_2$

Zhenkun Yuan, Yihuang Xiong, and Geoffroy Hautier*

*Thayer School of Engineering, Dartmouth College, Hanover, New Hampshire 03755, USA*

Email: geoffroy.hautier@dartmouth.edu

**Abstract**

Zinc phosphide ($Zn_3P_2$) has had a long history of scientific interest largely because of its potential for earth-abundant photovoltaics. To realize high-efficiency $Zn_3P_2$ solar cells, it is critical to understand and control point defects in this material. Using hybrid functional calculations, we assess the energetics and electronic behavior of intrinsic point defects and hydrogen impurities in $Zn_3P_2$. All intrinsic defects are found to act as compensating centers in *p*-type $Zn_3P_2$ and have deep levels in the band gap, except for zinc vacancies which are shallow acceptors and can act as a source of doping. Our work highlights that zinc vacancies rather than phosphorus interstitials are likely to be the main source of *p*-type doping in as-grown $Zn_3P_2$. We also show that Zn-poor and P-rich growth conditions, which are usually used for enhancing *p*-type conductivity of $Zn_3P_2$, will facilitate the formation of certain deep-level defects ($P_{Zn}$ and $P_i$) which might be detrimental to solar cell efficiency. For hydrogen impurities, which are frequently present in the growth environment of $Zn_3P_2$, we study interstitial hydrogen and hydrogen complexes with vacancies. The results suggest small but beneficial effects of hydrogen on the electrical properties of $Zn_3P_2$.

## 1. Introduction

The II-V semiconductor zinc phosphide ($Zn_3P_2$) is a promising light-absorbing material for cost-effective thin-film photovoltaics.[1, 2] It consists of earth-abundant elements and shows attractive optoelectronic properties as a solar absorber: a direct band gap of ~1.5 eV, a high absorption coefficient of over $10^4$ cm$^{-1}$,[3, 4] and minority-carrier diffusion lengths of up to a few micrometers.[5, 6] $Zn_3P_2$-based solar cells received considerable attention from the late 1970s to the 1990s. However, the efficiencies have been achieved as yet are far below the Shockley-Queisser limit (~33%).[7-12] The highest efficiency of $Zn_3P_2$ solar cells was reported in 1981 and has remained since then at just 5.96%.[8]

More recently, $Zn_3P_2$ solar cells have seen a resurgence of interest. Notably, researchers have developed fabrication methods for highly crystalline, reproducible $Zn_3P_2$ thin films on commercially available substrates which had long been a challenge.[13-15] The attainment of high-quality samples paves the way for addressing another persistent challenge: control of point defects and doping in bulk $Zn_3P_2$.[16, 17]

As-grown $Zn_3P_2$ crystals and thin films nearly always exhibit *p*-type conductivity, the origin of which is believed due to intrinsic point defects.[18] Because conductivity has been observed to vary with the phosphorus partial pressure in the growth or annealing environment, it has been commonly attributed to phosphorus interstitials ($P_i$).[12, 18-21] Despite no direct experimental observations of phosphorus interstitials or their involvement in *p*-type doping, this assignment has been supported by an early first-principles investigation based on semilocal density-functional theory (DFT), which has suggested that phosphorus interstitials are the most prevalent acceptor species in $Zn_3P_2$.[22] However, a subsequent first-principles study, using more accurate hybrid functional, has shown that $P_i$ is a deep acceptor and has much higher formation energy than another acceptor, the zinc vacancy ($V_{Zn}$).[23] In order to clarify these conflicting results, recently Stutz *et al.* and Paul *et al.* studied the effect of compositional stoichiometry variations on the structural, electrical, and optical properties of monocrystalline $Zn_3P_2$ thin films, yet a conclusive identification of the source for the *p*-type conductivity, whether it is due to $P_i$ or $V_{Zn}$, has remained elusive.[24, 25] On the other hand, less attention has been devoted to identifying deep-level defects in $Zn_3P_2$, in spite of accumulating experimental evidence for presence of deep levels in the band gap of $Zn_3P_2$.[26-28] Deep-level defects may act as nonradiative carrier recombination centers which would limit the efficiency of $Zn_3P_2$ solar cells.

In addition to intrinsic defects, there is strong experimental evidence that hydrogen impurities are likely present in $Zn_3P_2$ samples: growth of $Zn_3P_2$ thin films is conventionally carried out in $H_2$ atmosphere and/or $PH_3$ gas using techniques such as physical vapor transport,[4, 29] chemical vapor deposition[30, 31] and metal-organic chemical vapor deposition,[32-35] ionized cluster beam deposition,[36] and RF sputtering.[37, 38] Sometimes the samples were also annealed in $H_2$ gas following growth.[33, 37, 39-41] It is known that unintentional hydrogen doping in oxides and III-V nitrides has a strong impact on the electrical properties of these materials.[42-44] Therefore, in order to avoid uncontrolled influence of hydrogen impurities, it is important to understand the behavior of hydrogen in $Zn_3P_2$.

Using first-principles calculations with a hybrid functional, we study intrinsic point defects and hydrogen impurities in $Zn_3P_2$: zinc and phosphorus vacancies ($V_{Zn}$ and $V_P$), interstitials ($Zn_i$ and $P_i$), and antisites ($Zn_P$ and $P_{Zn}$); for hydrogen impurities, hydrogen interstitial ($H_i$) and hydrogen-vacancy complexes are investigated. The present work assesses the energetics and electronic behavior of the defects based on their calculated formation energies and charge transition levels. Comparing our results with experiments and previous calculations, we address the open questions about the shallow or deep nature of the defects in Zn$_3$P$_2$. Our work clarifies the likely defects leading to *p*-type doping in Zn$_3$P$_2$ and the possible deep-level defects which may act as nonradiative recombination centers. It also elucidates the role of hydrogen in doping of Zn$_3$P$_2$.

## 2. Intrinsic point defects

We have performed hybrid functional calculations for the intrinsic defects $V_{\text{Zn}}$, $V_{\text{P}}$, $\text{Zn}_i$, $\text{P}_i$, $\text{Zn}_{\text{P}}$, and $\text{P}_{\text{Zn}}$ in Zn$_3$P$_2$. Fig. 1 shows the calculated formation energies for these defects, under Zn-poor (P-rich) and Zn-rich (P-poor) chemical-potential conditions. Details about the calculation methods and the chemical potentials of Zn and P can be found in the Methods section. We find, in Fig. S1 of the Supplementary Information, a rather small chemical-potential region for which Zn$_3$P$_2$ is thermodynamically stable, due to strong phase competition from ZnP$_2$ (another stable crystalline phase in the Zn-P system).[45] Note that the terms "rich" and "poor" used in Fig. 1 and Fig. S1 do not represent the actual compositional stoichiometry of experimental samples but rather they correspond to the endpoints of the chemical-potential region that stabilizes compositionally stoichiometric Zn$_3$P$_2$, so the results presented in Fig. 1 are for the intrinsic defects in (nearly) stoichiometric samples.

As shown in Fig. 1, the $V_{\text{Zn}}$ and $\text{Zn}_i$ behave exclusively as an acceptor and as a donor, respectively. In contrast, all other intrinsic defects, including $V_{\text{P}}$, $\text{P}_i$, $\text{Zn}_{\text{P}}$, and $\text{P}_{\text{Zn}}$, are amphoteric: they exist in donor states when the Fermi level is low in the band gap while in acceptor states when the Fermi level is high in the band gap. The amphoteric behavior of these defects is related to the unique ability for phosphorus to occur in multiple oxidation states from $\text{P}^{3-}$ to $\text{P}^{5+}$.

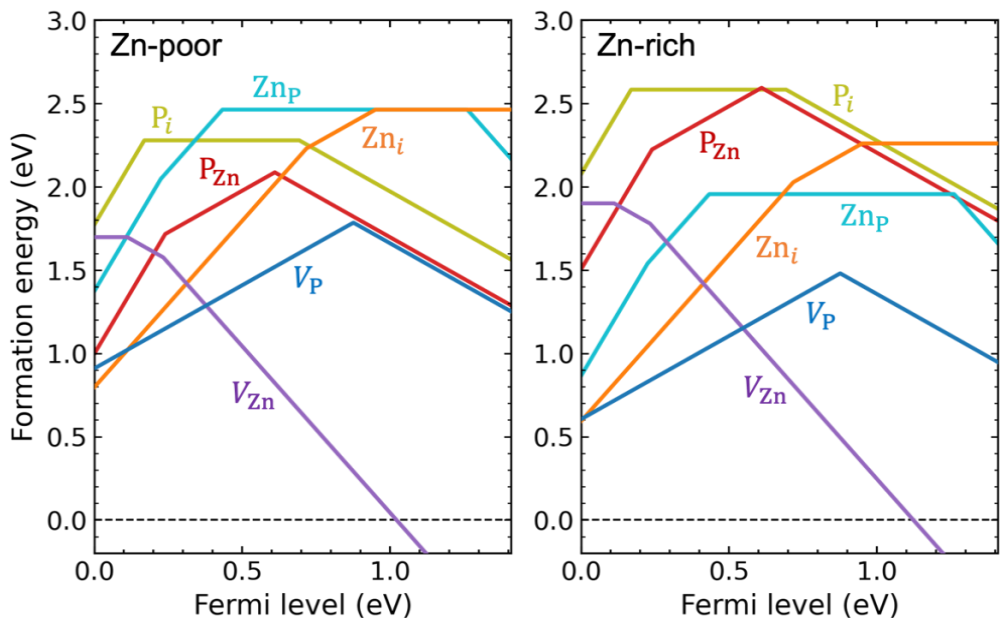

**Fig. 1** Formation energies as a function of Fermi level for intrinsic point defects in Zn$_3$P$_2$, under Zn-poor/P-rich (left) and Zn-rich/P-poor (right) conditions. The zero of the Fermi level is set at the VBM and the upper bound corresponds to the conduction-band minimum (CBM). For the same Fermi-level value, only formation energy for the most stable charge state is shown, i.e., $V_{\text{Zn}}$: 0, −, 2 −; $V_{\text{P}}$: +, −; $\text{Zn}_i$: 2 +, +, 0; $\text{P}_i$: 3 +, 0, −; $\text{Zn}_{\text{P}}$: 3 +, 2 +, 0, 2 −; $\text{P}_{\text{Zn}}$: 3 +,

+, −. Kinks in each curve indicate transitions between different charge states (Values of the thermodynamic transition levels are also listed in Table S1 of the Supplementary Information).

We find that the $V_{Zn}$ behaves as a shallow acceptor and is the only intrinsic defect that can act as a source of *p*-type doping in $Zn_3P_2$. The $V_{Zn}$ gives rise to two acceptor levels in the band gap: a shallow $(0/−)$ transition level at 0.11 eV and a relatively deep $(−/2−)$ transition level at 0.23 eV above the valence band (see Fig. 1). For reference, the calculated single-particle defect states of $V_{Zn}$ can be found in Fig. S2 of the Supplementary Information. The formation energy of $V_{Zn}$ is relatively high for Fermi-level positions close to the valence-band maximum (VBM), even under Zn-poor/P-rich condition.

Contrary to the commonly accepted assumption that phosphorus interstitials are the cause of *p*-type conductivity in as-grown $Zn_3P_2$,[12, 18-21] we find that phosphorus interstitials cannot possibly contribute to *p*-type doping. First and foremost, $P_i$ gives rise to a very deep acceptor level $\epsilon(0/−) = 0.69$ eV (see Fig. 1); thus, it can hardly provide holes to the valence band through thermal excitation. Second, for most Fermi-level positions in the lower part of the band gap, $P_i$ will exist in the neutral charge state (being electrically inactive). Even when the Fermi level is close to the valence band, $P_i$ can also be stable in the $+3$ charge state, with a $(3+/0)$ transition level at 0.17 eV above the VBM. The existence of this positive charge state, which removes three holes from the valence band, suggests that phosphorus interstitials could act as compensating centers in *p*-type $Zn_3P_2$ (though this happens for Fermi level below 0.17 eV). Third, $P_i$ has a high formation energy and is therefore not expected to form in significant concentrations in (nearly) stoichiometric $Zn_3P_2$ samples.

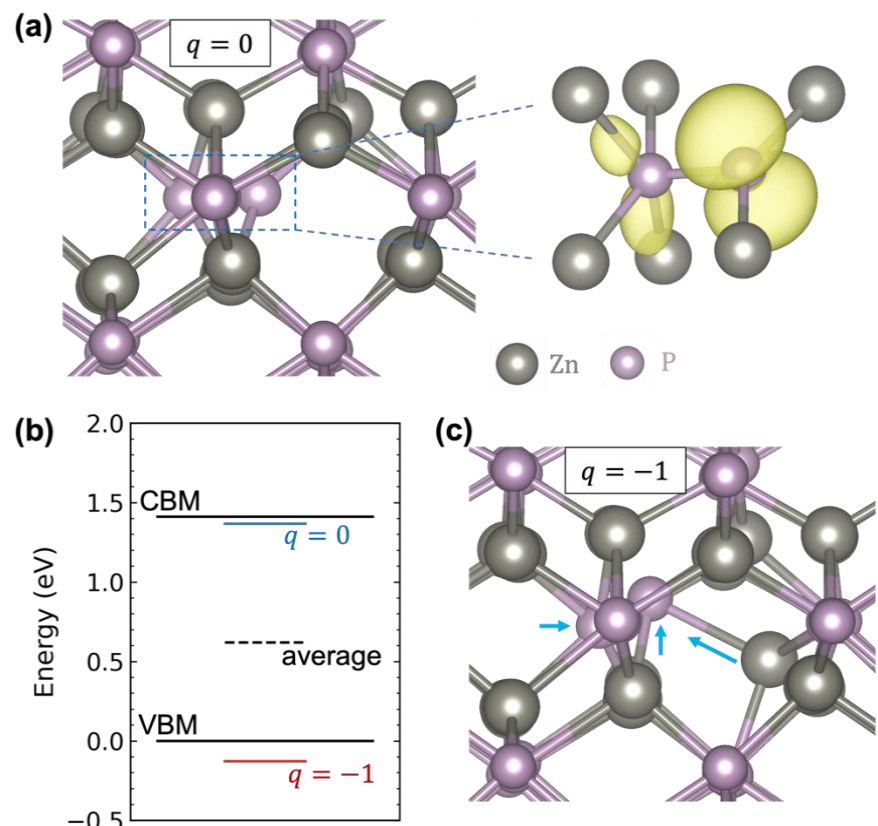

**Fig. 2** (a) Local atomic geometry and spin density (isosurface at 10% of the maximum in yellow) associated with the localized hole state of the $P_i$ in the neutral charge state ($q = 0$). The $P_i$ adopts a split-interstitial configuration in which the interstitial P shares a lattice site with a lattice P atom. (b) Single-particle energy levels of the localized hole state of the $P_i$, before and after being occupied. The energy of the VBM is set to zero. (c) Local atomic geometry of the $P_i$ in the $q = -1$ charge state, with the arrows indicating the local atomic relaxations.

Fig. 2a shows that in the neutral charge state, the $P_i$ adopts a split-interstitial configuration in which the interstitial P shares a lattice site with one of the lattice P atoms, consistent with previous calculations.[22, 23] Similar split-interstitial configuration has been found for the nitrogen interstitial in GaN.[46] The plotted spin density shows that the neutral $P_i$ is characterized by a highly localized hole state, with most of the charge located on the interstitial P atom and the bonded lattice P atom. The calculated single-particle energy level of the hole state is located well above the VBM (more specifically, right below the CBM), as indicated by the blue line in Fig. 2b. When the $P_i$ becomes negatively charged, there is a large relaxation of the interstitial and the surrounding atoms, yet the split-interstitial configuration is largely kept, as shown in Fig. 2c. As a result, the single-particle defect level is pushed down to below the VBM, as indicated by the red line in Fig. 2b. Interestingly, as indicated by the black dashed line in Fig. 2b, the average of the two single-particle defect levels is 0.62 eV above the VBM, which gives a good estimate of the $(0/-)$ transition level (0.69 eV).

All other intrinsic defects, including $V_P$, $Zn_i$, $Zn_P$, and $P_{Zn}$, have deep transition levels in the band gap, as seen in Fig. 1. The $Zn_i$ is a double donor with two deep levels: a $(+/0)$ transition level at 0.95 eV and a $(2+/+)$ transition level at 0.72 eV above the VBM. For $V_P$, the formation energies of $V_P^+$ and $V_P^-$ intersect at 0.88 eV above the VBM, and $V_P^0$ is energetically unstable compared to either $V_P^+$ or $V_P^-$ over the entire range of Fermi-level positions in the band gap; this is characteristic of a negative-$U$ behavior. Similar to $V_P$, the $P_{Zn}$ forms a deep transition level between $+1$ and $-1$ charge states at 0.61 eV above the VBM. The $Zn_P$ has a deep $(2+/0)$ transition level (also a negative-$U$ transition) at 0.43 eV above the VBM. Besides these deep levels, the $P_{Zn}$ $(3+/+)$ transition level is close to the VBM, and the $Zn_P$ $(3+/2+)$ and $(0/2-)$ transition levels are close to the VBM and CBM, respectively.

Similar to the $P_i$, the $V_P$, $Zn_P$, and $P_{Zn}$ will contribute to compensation in $p$-type $Zn_3P_2$, due to their amphoteric behavior. As indicated in Fig. 1, the $V_P$ and $Zn_i$ are the main compensating intrinsic defects in $p$-type $Zn_3P_2$. In addition, depending on growth conditions and on the $p$-type doping level (i.e., the Fermi-level position), the formation energy of $Zn_P$ and $P_{Zn}$ can be as low as ~ 1 eV, suggesting that both defects can form in significant concentrations and thus also be important compensating centers. From Fig. 1 we can see that compensation of $p$-type doping is reduced under Zn-poor/P-rich condition where the $V_P$ and $Zn_i$ have higher formation energy (though the formation energy of $P_{Zn}$ is lowered under this condition).

## 3. Comparison with experiments and discussion

The type of possible doping or dopability of a material are largely controlled by compensation. Hole-killer (electron-killer) defects can prevent $p$-type ($n$-type) doping.[47, 48] Our first-principles results indicate that $Zn_3P_2$ clearly favors $p$-type doping, because no hole-killer defects will pin the Fermi level while the electron-killer $V_{Zn}$ will pin the Fermi level and prevent $n$-type doping even in the presence of shallow extrinsic donors. The $p$-type doping of $Zn_3P_2$ is favored under Zn-poor/P-rich conditions, where both the formation energy of $V_{Zn}$ acceptors and the compensation will be reduced. Our finding agrees with the experimentally observed $p$-type nature of $Zn_3P_2$.[12, 18-21]

Our results suggest that $V_{Zn}$ (but not $P_i$) is a source of *p*-type conductivity in as-grown Zn3P2. This disagrees with the previous reports that *p*-type conductivity in Zn3P2 is due to $P_i$ which has been based on experimental observations that conductivity of Zn3P2 samples varies with the phosphorus partial pressure in the growth or annealing environment.[12, 18-21] Here we take a closer look at Refs. 18 and 19. In Ref. 18, Zn3P2 samples were postgrowth annealed over a range of equilibrium vapor compositions, and conductivity was observed to be increased when changing from Zn-rich annealing conditions (zinc vapor) to P-rich annealing conditions (phosphorus vapor). In Ref. 19, electrical measurements were performed on a set of samples with the P/(Zn + P) ratio varying from 0.39 (which corresponds to a P-poor sample) to 0.4 (which corresponds to a stoichiometric sample), showing increased conductivity. For both studies, the observed conductivity increase can be explained by an enhancement of the $V_{Zn}$ concentration and a reduction of the $V_P$ and $Zn_i$ concentrations. It should be noted that since Zn3P2 is a binary compound, a P-rich (P-poor) condition means a Zn-poor (Zn-rich) condition. We conclude that our results are in line with the experimental observations.

The more recent experimental work by Stutz *et al.*,[24] is also of interest. This work studied monocrystalline, nonstoichiometric P-rich Zn3P2 thin films and found increasing lattice expansion when the compositional stoichiometry varies from Zn2.98P2.02 (Zn/P=1.47) to Zn2.75P2.25 (Zn/P=1.22) to Zn2.67P2.33 (Zn/P=1.15). Stutz *et al.* interpreted their results in terms of the formation of $P_i$, and also suggested that many of the defects in their samples are in neutral charge state (based on the fact that in all three samples, the compositionally induced defect density is about five orders of magnitude higher than carrier density). Our results support the analysis of Stutz *et al.*, whereas further indicate that it is necessary to account for the $P_{Zn}$ in understanding the defects in non-stoichiometric P-rich Zn3P2 samples; that is, an excess of P leads to $P_{Zn}$ and $P_i$ and facilitates the formation of $V_{Zn}$. In this case, the $V_{Zn}$ acceptors are expected to be heavily compensated by $P_{Zn}$, and most $P_i$ defects will be in the neutral charge state as discussed above. Within this picture, non-stoichiometric P-rich Zn3P2 samples should be highly compensated.

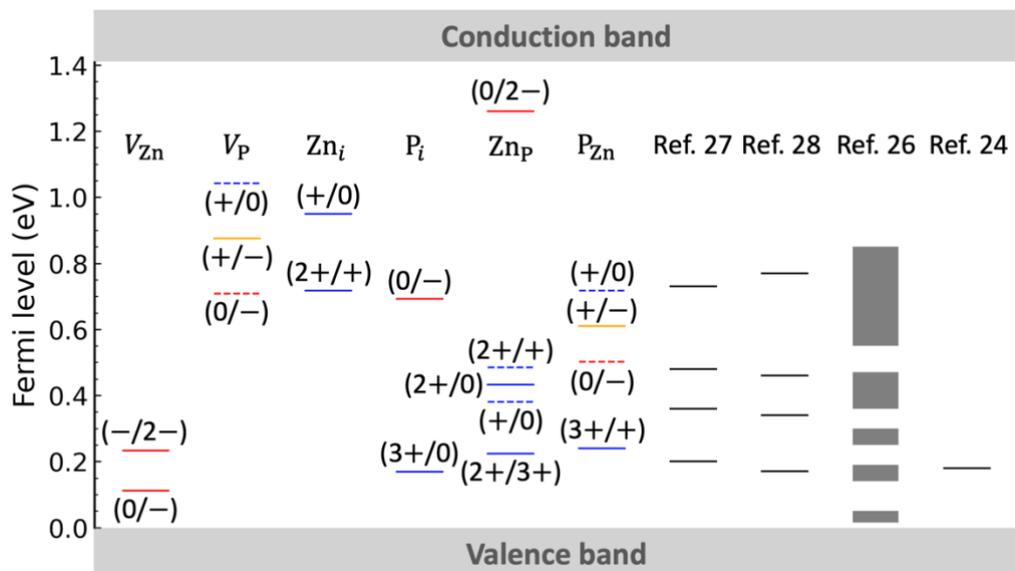

**Fig. 3** Comparison of our calculated thermodynamic transition levels (between different charge states $q$ and $q'$) with experimental values reported in the literature. The red, blue, and orange solid lines indicate the calculated acceptor, donor, and $(+/-)$ transition levels, respectively. The dashed lines indicate the relevant transition levels involving thermodynamically unstable

charge states. Experimental data are taken from Refs. 24, 26-28; they have been reported as acceptor levels or hole-trap levels in the original work.

Fig. 3 shows a comparison of our calculated positions of charge transition levels with experimental values obtained by deep-level transient spectroscopy (DLTS; on $Zn_3P_2$ polycrystals),[27, 28] electrical transport measurements (on $Zn_3P_2$ single crystals),[26] and photoluminescence spectroscopy (on monocrystalline, nonstoichiometric P-rich $Zn_3P_2$ thin films).[24] Note that the experimental results have been classified as acceptor levels or hole-trap levels in the original work. In Fig. 3 we also plot the relevant transition levels involving thermodynamically unstable charge states: the $(+/0)$ and $(0/-)$ transition levels of both $V_P$ and $P_{Zn}$, and the $(2+/+)$ and $(+/0)$ transition levels of $Zn_P$ (For more information on the charge-state transitions we refer to Fig. S3 of the Supplementary Information); such levels could still be detected by DLTS.[49] In the following, we attempt to identify the origin of the experimentally observed defect levels. This assignment is tentative and based purely on the defect energy levels. It does not take into account defect complexes or extrinsic impurities which could be present in the experimental samples.

First, we assign the acceptor level lying at $\sim 0.14 - 0.2$ eV, commonly observed in the experiments, to the $(-/2-)$ transition level of $V_{Zn}$. In Ref. 26 and another electrical transport experiment,[50] it has been found that in $Zn_3P_2$ samples with high hole density ($\sim 10^{17}$ cm$^{-3}$), there is only one important acceptor level lying at $\sim 0.05$ eV. This shallow acceptor level is assigned tentatively to the $(0/-)$ transition level of $V_{Zn}$. Second, for the experimentally observed hole-trap levels, our speculations are as follows. The level lying at $0.25 - 0.30$ eV (observed by electrical transport measurements) is assigned tentatively to the $P_{Zn}$ $(3+/+)$ transition level. The level lying at $\sim 0.35$ eV (detected by DLTS) is identified tentatively as the $Zn_P$ $(+/0)$ transition level. The level lying at $\sim 0.46$ eV (DLTS) may be assigned to $Zn_P$ $(2+/0)$ or $(2+/+)$ or $P_{Zn}$ $(0/-)$ transition levels. The level lying at $\sim 0.75$ eV (DLTS) may be related to the $V_P$ $(0/-)$, $Zn_i$ $(2+/+)$, $P_i$ $(0/-)$, or $P_{Zn}$ $(+/0)$ transitions.

Deep levels in a solar absorber could cause nonradiative recombination of photo-generated carriers and hence be detrimental to the device performance.[51-54] Our identification of the intrinsic defects having deep levels in $Zn_3P_2$ will help optimize the absorber by control of these defects. However, suppressing the formation of the deep-level defects is likely to be challenging for $Zn_3P_2$. For instance, while Zn-poor/P-rich conditions are needed to enhance p-type doping, such conditions will facilitate the formation of $P_{Zn}$ and $P_i$ which possess deep levels. This implies that good photovoltaic performance is not guaranteed for nonstoichiometric P-rich $Zn_3P_2$ samples which have become the focus of recent experimental studies.[17, 24, 25] This analysis would however require a systematic study of the nonradiative carrier capture coefficients (cross sections) which could vary considerably between different deep levels,[55-59] which is beyond the scope of the present work.

## 4. Comparison with previous calculations

The first DFT study of intrinsic defects in $Zn_3P_2$ by Demers *et al.* used a semilocal functional.[22] One of the main conclusions of Ref. 22 is that phosphorus interstitials are the most prevalent acceptors. Because of the inherent limitations of semilocal functionals in predicting band gaps for semiconductors,[60-62] Demers *et al.* corrected the band gap of $Zn_3P_2$ using an extrapolation method that is often difficult to justify.[60] In light of this deficiency, we believe that the calculated defect energetics in Ref. 22 suffer from considerable uncertainties.

On the other hand, Yin *et al.* have reported hybrid-functional calculations for intrinsic defects in Zn$_3$P$_2$.[23] Both our results and those of Yin *et al.* agree that: (i) the $V_{Zn}$ behaves as a shallow acceptor; (ii) the P$_i$ gives rise to a deep acceptor level, and is energetically less favorable than $V_{Zn}$. However, some notable differences exist. As listed in Table S2 of the Supplementary Information, the charge transition levels reported by Yin *et al.* are systematically shallower than ours. For instance, the $(0/-)$ transition level of P$_i$ is 0.30 eV in Ref. 23 while it is 0.69 eV in the present work. In addition, Yin *et al.* have shown that $V_P$ and P$_i$ behave exclusively as a donor and as an acceptor, respectively, while we find amphoteric behavior of both defects. We attribute these discrepancies to the apparent lack of spin polarization and proper supercell-size corrections in the defect calculations in Ref. 23.

## 5. Effects of hydrogen on doping and defect passivation

We now turn to hydrogen impurities in Zn$_3$P$_2$. We have considered hydrogen interstitial (H$_i$) and hydrogen in zinc and phosphorus vacancies, and performed a careful search for their lowest-energy configurations. It is seen in Fig. S4 of the Supplementary Information that H$_i^0$ and H$_i^-$ are most stable at the tetrahedral site and H$_i^+$ is most stable at the bond-center site. For a single H in the zinc vacancy, we initially incorporated it on the substitutional Zn site but found that it moves off to bind with one of the four surrounding P atoms; it is thus regarded as a H + $V_{Zn}$ complex rather than a substitutional H$_{Zn}$. The $V_{Zn}$ can trap a second H to form a 2H + $V_{Zn}$ complex in which each H binds with one P atom. For a single H in the phosphorus vacancy, the H is also located off-center but binds with two Zn atoms, resulting in a H + $V_P$ complex.

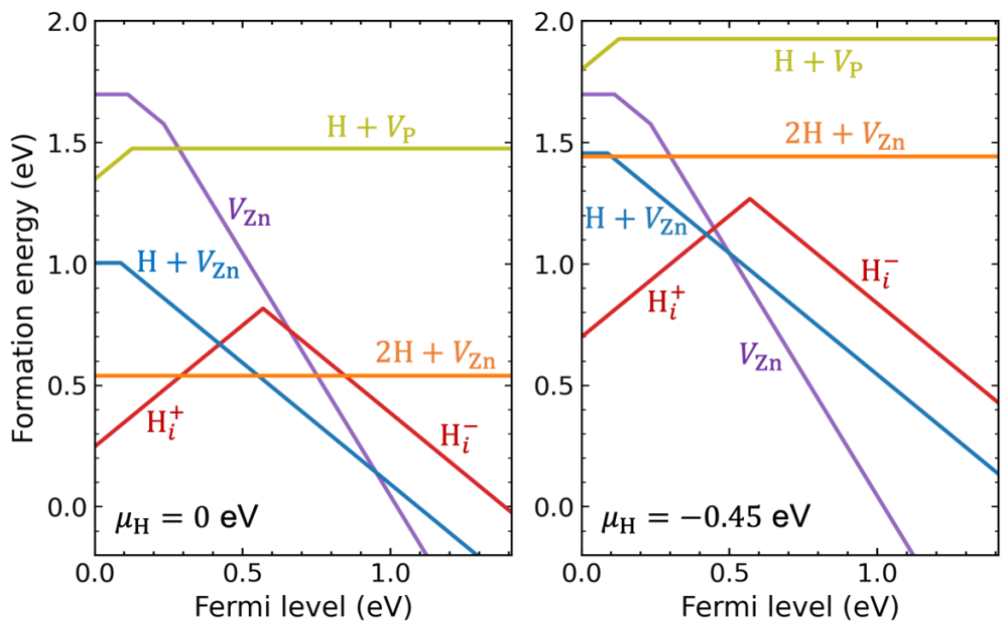

**Fig. 4** Formation energies as a function of Fermi level for hydrogen interstitial (H$_i$) and hydrogen-vacancy complexes, including the H + $V_{Zn}$, 2H + $V_{Zn}$, and H + $V_P$, in Zn$_3$P$_2$, under Zn-poor/P-rich condition and at two different chemical potentials of H ($\mu_H$): the H-rich limit $\mu_H = 0$ eV (left panel) and a lower $\mu_H$ at $-0.45$ eV (right panel). For comparison the formation energy of $V_{Zn}$ is also depicted.

Fig. 4 plots the calculated formation energies for the H-related point defects, under Zn-poor/P-rich condition (as in Fig. 1) and at two different chemical potentials of H ($\mu_H$): (i) $\mu_H = 0$, i.e., the H-rich limit; (ii) a lower $\mu_H$ of $-0.45$ eV (This value is a more realistic choice of $\mu_H$; it can be obtained through equilibrium with H$_2$ gas at 400 °C and partial pressure of 0.01 atm in the ideal-gas model;[63, 64] see the Methods section). Let us first analyze the electronic behavior of the H-related defects. As can be seen, H$_i$ is amphoteric with a $(+/-)$ transition level in the band gap, as commonly found for hydrogen interstitial in semiconductors.[65] Since the $(+/-)$ transition level is 0.57 eV above the VBM, H$_i$ acts as a compensating center in *p*-type Zn$_3$P$_2$. Further, hydrogen passivates the dangling bonds of the vacancies and hence reduce their charge states. The H + $V_{Zn}$ complex is a single acceptor with a $(0/-)$ transition level 0.09 eV above the VBM, which is slightly shallower than that of isolated $V_{Zn}$ (0.11 eV). The 2H + $V_{Zn}$ complex has no gap states and does not introduce any defect level in the band gap. The H + $V_P$ complex is a single donor with a $(+/0)$ transition level just 0.13 eV above the VBM.

Under the H-rich limit (left panel of Fig. 4), we see that the H$_i^+$, H + $V_{Zn}$, and 2H + $V_{Zn}$ have low formation energy. In contrast, the H + $V_P$ has relatively high formation energy. Here and in the following, our implicit assumption is that we are considering *p*-type Zn$_3$P$_2$. Moving to the lower $\mu_H$ (right panel of Fig. 4), the formation energies of the H-related defects are increased, especially for the 2H + $V_{Zn}$ complex. In addition, the formation energy of H + $V_{Zn}$ and 2H + $V_{Zn}$ becomes comparable to that of isolated $V_{Zn}$. Irrespective of $\mu_H$, the binding energy of the [H + $V_{Zn}$]$^-$ (2H + $V_{Zn}$) with respect to isolated $V_{Zn}^{2-}$ and H$_i^+$ is found to be 1.2 eV (2.0 eV). This is quite a large value, indicating that hydrogen interacts strongly with zinc vacancies resulting in stable complexes. The binding energy of the H + $V_P$ with respect to isolated $V_P^-$ and H$_i^+$ is also found to be large (1.43 eV). This suggests that the H + $V_P$ complex would be stable if formed, despite having a relatively high formation energy. We also take note of the fact that interstitial hydrogen is expected to be quite mobile in crystalline semiconductors.[66-68]

It is thus found that when incorporated into Zn$_3$P$_2$, hydrogen is likely to form H$_i^+$ and stable complexes with the $V_{Zn}$ acceptors. Because, compared with isolated $V_{Zn}$, the H + $V_{Zn}$ has shallower acceptor level and lower formation energies (under H-rich conditions), hydrogen incorporation may enhance *p*-type doping of Zn$_3$P$_2$. As shown in Fig. S5 of the Supplementary Information, the calculated bulk hole density increases as $\mu_H$ approaches zero (the H-rich limit). Nevertheless, the enhancement of *p*-type doping by H + $V_{Zn}$ is found to be small, because of the compensation by H$_i^+$ and the competition from 2H + $V_{Zn}$ (which is electrically inactive).

We are aware of the experimental work of Wang *et al*.[39] and Bube[40], which show increased bulk hole density (by about an order of magnitude) in Zn$_3$P$_2$ samples as a result of high-temperature annealing in H$_2$ gas. Yet, Bube[40] noted a depletion of holes in the surface region of the samples after the annealing. Suda *et al*.[37] reported that annealing Zn$_3$P$_2$ films in H$_2$ gas reduces the resistivity by an order of magnitude. In another study, Suda *et al*.[27] found that hydrogen plasma treatment of Zn$_3$P$_2$ removes a native hole-trap level at 0.2 eV above the VBM, which can be explained by passivation of $V_{Zn}$. Suda *et al*. also observed hydrogen passivation of deep hole-trap levels, and found that excessive hydrogenation creates a new hole-trap level at 0.13 eV above the VBM, which may be explained by formation of the H + $V_P$ complex. For hydrogen incorporation during the deposition of Zn$_3$P$_2$ films, Kakishita *et al*.[36] found only a passivation effect of hydrogen, but not change in conductivity.

## 6. Conclusions

In conclusion, we have studied intrinsic point defects and hydrogen impurities in Zn₃P₂ using first-principles hybrid functional calculations. Among the intrinsic defects, only zinc vacancies are shallow acceptors and can act as a source of *p*-type doping. All other intrinsic defects, including the $V_\text{P}$, $\text{Zn}_i$, $\text{P}_i$, $\text{Zn}_\text{P}$, and $\text{P}_\text{Zn}$, will contribute to compensation in *p*-type Zn₃P₂. All of these defects are amphoteric (except for the $\text{Zn}_i$ which is exclusively a donor), and they all have deep levels in the band gap. Control of the intrinsic defects by adjusting the growth conditions to enhance *p*-type doping may be challenging for Zn₃P₂ if the concentration of the deep-level defects has to be kept low. For instance, while Zn-poor/P-rich growth or annealing conditions are needed to enhance the concentration of $V_\text{Zn}$ acceptors, such conditions will facilitate the formation of the deep $\text{P}_\text{Zn}$ and $\text{P}_i$ defects. Hydrogen impurities in Zn₃P₂ are likely to form $\text{H}_i^+$ and stable complexes with the $V_\text{Zn}$ acceptors. The H + $V_\text{Zn}$ complex is a shallow acceptor which can contribute to doping, while the $\text{H}_i^+$ will lead to compensation of holes and the 2H + $V_\text{Zn}$ complex is electrically inactive. Due to the counteracting behavior of the H-related defects, the impact of hydrogen on *p*-type doping of Zn₃P₂ is predicted to be small.

## 7. Methods

All the DFT calculations were performed using the projector augmented wave (PAW) pseudopotential method and hybrid functional of Heyd–Scuseria–Ernzerhof (HSE) as implemented in the VASP code.[69-71] The wave functions were expanded in a plane-wave basis set with an energy cutoff of 400 eV. The HSE mixing parameter ($\alpha$) was set to 0.32, which yields a band gap $E_\text{g} = 1.41$ eV and lattice parameters $a = 8.076$ Å and $c = 11.388$ Å for the Zn₃P₂ unit cell, in agreement with the experimental values ($E_\text{g} \sim 1.5$ eV, and lattice parameters $a = 8.09$ Å and $c = 11.45$ Å).[4, 72] A 2 × 2 × 2 supercell (which contains 320 atoms when it is defect free) was used for point-defect simulations, with a $\Gamma$ point for the Brillouin-zone sampling. All atomic coordinates in supercells containing a point defect were fully relaxed until the residual atomic forces become less than 0.01 eV/Å. Spin polarization was explicitly considered in all the defect calculations.

Defect formation energies and thermodynamic charge transition levels were calculated using the standard first-principles formalism as detailed in Ref. 61 and the PyCDT code.[73] For charged defects, electrostatic correction and potential alignment were applied using the experimental static dielectric constant ($\varepsilon_0 = 11$).[74-76] Since the formation energy of a defect depends on the chemical potential of the elements involved in creating the defect, we have determined the chemical potential of Zn ($\mu_\text{Zn}$) and the chemical potential of P ($\mu_\text{P}$) according to the following four relations:

$$\mu_\text{Zn} < 0,$$
$$\mu_\text{P} < 0,$$
$$\mu_\text{Zn} + 2\mu_\text{P} < \Delta H_\text{f}(\text{ZnP}_2),$$
$$3\mu_\text{Zn} + 2\mu_\text{P} = \Delta H_\text{f}(\text{Zn}_3\text{P}_2),$$

where: (i) the first relation means $\mu_\text{Zn}$ (referenced to the energy of Zn in bulk zinc) should be less than 0 eV to avoid formation of bulk zinc phase; (ii) the second relation means $\mu_\text{P}$ (referenced to the energy of P in bulk phosphorus) should be less than 0 eV to avoid formation of bulk phosphorus phase; (iii) the third relation means $\mu_\text{Zn}$ and $\mu_\text{P}$ are bound such that the secondary phase ZnP$_2$ will not form; and (iv) the last relation ensures thermodynamic stability of (compositionally stoichiometric) Zn$_3$P$_2$. Using the calculated heats of formation (per unit formula): $\Delta H_f(\text{ZnP}_2) = -0.91$ eV and $\Delta H_f(\text{Zn}_3\text{P}_2) = -1.31$ eV, the allowed chemical-potential region of Zn$_3$P$_2$ is determined and plotted in Fig. S1. Due to strong phase competition from ZnP$_2$, the stable region of Zn$_3$P$_2$ is quite small. Two different chemical-potential points of the stable region: A ($\mu_{\text{Zn}} = -0.203$ eV, $\mu_{\text{P}} = -0.354$ eV) and B ($\mu_{\text{Zn}} = 0.0$ eV, $\mu_{\text{P}} = -0.658$ eV), which represent Zn-poor (P-rich) and Zn-rich (P-poor) equilibrium growth conditions, respectively, were used in computing the defect formation energies.

The formation energy of the H-related point defects in Zn$_3$P$_2$ also depends on the chemical potential of H ($\mu_{\text{H}}$). In the H-rich limit, $\mu_{\text{H}} = 0$ eV [referenced to half the energy $E(\text{H}_2)$ of an H$_2$ molecule at $T = 0$ K]. We have also considered the familiar situation of annealing in H$_2$ gas at temperature $T$ and partial pressure $p$, where $\mu_{\text{H}}$ is determined through equilibrium with H$_2$ gas. In this case, $\mu_{\text{H}}$ is a well-established function of $T$ and $p$:

$$\mu_{\text{H}}(T,p) = \frac{1}{2}E(\text{H}_2) + \mu_{\text{H}}^{\text{H}_2 \text{ gas}}(T,p),$$

The $\mu_{\text{H}}^{\text{H}_2 \text{ gas}}(T,p)$ for moderate $T$ and $p$ can be obtained by an analytical expression arising from the ideal-gas model [see e.g., Eq. (21) of Ref. 63]. In the main text, we considered $T = 400$ °C and $p = 0.01$ atm, which gives $\mu_{\text{H}} = -0.45$ eV.

Finally, in order to check whether our choice of $\alpha = 0.32$ for the HSE mixing parameter describes correctly the localized hole state of P$_i^0$, the fulfillment of the generalized Koopmans' theorem has been examined.[77, 78] We find the non-Koopmans energy for P$_i^0$, defined as $E_{NK} = \epsilon(N) - [E(N+1) - E(N)]$, to be small (0.12 eV). Here, $\epsilon(N)$ is the single-particle energy level of the localized hole state of P$_i^0$, and $E(N+1) - E(N)$ is the total-energy difference between P$_i^-$ (with atomic positions fixed to those of P$_i^0$) and P$_i^0$. The electrostatic finite-size correction was only applied to the total energy of P$_i^-$, and $E_{NK}$ may still contain a small finite-size error. We conclude that the self-interaction error is relatively small in the HSE description of the localized hole state of P$_i^0$. We have also used the standard HSE06 functional ($\alpha = 0.25$) to calculate the $(0/-)$ transition level of the P$_i$, starting from the HSE06-calculated lattice parameters. The HSE06 calculations also find a split-interstitial configuration of P$_i^0$, and yield a $(0/-)$ transition level 0.54 eV above the VBM.

**Conflicts of interest**

There are no conflicts of interest to declare.

**Acknowledgements**

This work was supported by the U.S. Department of Energy, Office of Science, Basic Energy Sciences under Award Number DE-SC0023509. This research used resources of the National Energy Research Scientific Computing Center (NERSC), a DOE Office of Science User Facility supported by the Office of Science of the U.S. Department of Energy under Contract No. DE-AC02-05CH11231 using NERSC award BES-ERCAP0020966.